\documentclass[copyright,creativecommons]{eptcs}

\providecommand{\event}{FICS 2024} 

\usepackage[LGR,T1]{fontenc}
\usepackage{newtxtext,newtxmath}

\usepackage[utf8]{inputenc}
\usepackage{underscore}
\usepackage{alphabeta}

\usepackage[english]{babel}
\usepackage{xspace}
\usepackage{csquotes}

\usepackage{mathtools}
\usepackage{bm}
\usepackage[bbgreekl]{mathbbol}

\usepackage{xparse}
\usepackage{etoolbox}
\usepackage{array}

\usepackage{xcolor}
	\definecolor{c1}{Hsb}{210, 0.9, 0.7}
	\definecolor{c2}{Hsb}{160, 0.9, 0.7}
	\colorlet{c0}{black}
	\newcommand{\co}[1]{\color{c#1}}

\usepackage{cleveref}
\crefname{diagram}{diagram}{diagrams}

\RequirePackage{tikz}
	\usetikzlibrary{positioning}
	\usetikzlibrary{arrows.meta}
\usepackage{tikz-cd}
	\tikzcdset{arrow style=math font}

\RequirePackage{ebproof}
	\newcommand{\regle}[1]{(#1)}
	\ebproofset{ right label template={$\regle{\inserttext}$} }

\RequirePackage{float}
	\floatstyle{boxed}
	\restylefloat{table}
	\restylefloat{figure}
\RequirePackage{subcaption}
\usepackage{thmtools}
	\declaretheorem[name=Theorem]{thm}
	\declaretheorem[numberlike=thm, name=Lemma]{lem}
	\declaretheorem[numberlike=thm, name=Corollary]{cor}
	\declaretheorem[numberlike=thm, name=Definition]{defi}
	\declaretheorem[numberlike=thm, name=Notation]{nota}
	\declaretheorem[numberlike=thm, name=Example]{exa}

\newcommand{\symbf}[1]{\mathbf{#1}}
\newcommand{\symbb}[1]{\mathbb{#1}}
\newcommand{\symfrak}[1]{\mathfrak{#1}}
\let\transp=\relax

\usepackage{trimclip}

\definecolor{TODO}{RGB}{255,100,0}
\definecolor{R}{RGB}{244,7,110}
\definecolor{L}{RGB}{6,170,162}

\newcommand{\ie}{\emph{i.e.}\xspace}
\newcommand{\eg}{\emph{e.g.}\xspace}

\colorlet{c0}{black}
\definecolor{c3}{HTML}{AA6600}

\newcommand{\Bool}{\mathbb{B}}
\newcommand{\Nat}{\mathbb{N}}
\NewDocumentCommand \Reals 			{o}		{ \symbb{R}_{\IfValueT{#1}{#1}} }

\NewDocumentCommand \transp 		{mm}	{ \left( #1 \ #2 \right) }

\newcommand			\defeq				{ \coloneq }
\newcommand			\setcomp	[2]		{ \left\{\,#1\,\middle|\,#2\,\right\} }
\NewDocumentEnvironment{inductivedef}{s}{
		\newcommand{\sep}{\mathrel{|}}
		\IfBooleanTF{#1}{
			\begin{array}{ r !{\hspace{.5em}\ni\hspace{.5em}}
			r !{\hspace{.5em}\defeq\hspace{.5em}} l
			>{\qquad (}r<{)} }
		}{
			\begin{array}{ r !{\hspace{.5em}\defeq\hspace{.5em}} l
			>{\qquad (}r<{)} }
		}
	}{
		\end{array}
	}
\crefname{grammar}{grammar}{grammars}

\newcommand{\later}{\mathop{\blacktriangleright}}

\NewDocumentCommand \nf 	{om} 	{ 
	\mathrm{nf}_{ \IfValueT{#1}{#1} }\left(#2\right)
}

\NewDocumentCommand \SBTree		{o}		{
	\mathrm{SBTree}\IfValueT{#1}{_{\omega,#1}}
}

\NewDocumentCommand \btlt 		{s}		{ 
	\IfBooleanTF{#1}{\sqsubset}{\sqsubseteq}
}

\NewDocumentCommand \llist 		{o}		{%
	\mathsf{list}\IfValueT{#1}{(#1)}
}

\NewDocumentCommand \diagfunc 	{}		{ \Delta }

\renewcommand		\to					{ \rightarrow }

\NewDocumentCommand	\CC 		{}		{ \symbf{C} }

\NewDocumentCommand \SET 		{}		{ \symbf{Set} }
\NewDocumentCommand \NOM 		{}		{ \symbf{Nom} }

\NewDocumentCommand \id 		{o}		{
	\mathrm{id}\IfValueT{#1}{_{#1}}
}

\NewDocumentCommand \lfp 		{}		{ \bbmu }
\NewDocumentCommand \gfp 		{}		{ \bbnu }
\NewDocumentCommand \fold 		{o}		{ \mathsf{fold}\IfValueT{#1}{_{#1}} }
\NewDocumentCommand \unfold 	{o}		{ \mathsf{unfold}\IfValueT{#1}{_{#1}} }
\NewDocumentCommand \ialg 		{m} 	{ \fold[#1] }

\NewDocumentCommand \fv 		{}		{ \mathrm{fv} }

\NewDocumentCommand	\aeq		{}		{ \mathrel{=_\alpha} }
\NewDocumentCommand	\qbalpha	{}		{ /\mathord{=_\alpha} }

\NewDocumentCommand \arite		{}		{ \mathsf{ar} }
\NewDocumentCommand \opcons		{}		{ \mathsf{cons} }
\NewDocumentCommand \sabs 		{m}		{ #1. } 
\newcommand{\bs}{\textsc{bs}\xspace}
\newcommand{\mbs}{\textsc{mbs}\xspace}

\NewDocumentCommand \oplam		{}		{ \mathsf{lam} }
\NewDocumentCommand \opnode		{}		{ \mathsf{node} }
\NewDocumentCommand \opdig		{}		{ \mathsf{dig} }

\NewDocumentCommand \invar		{}		{ \mathsf{invar} }
\NewDocumentCommand \incons		{}		{ \mathsf{incons} }
\NewDocumentCommand \inl		{}		{ \mathsf{inl} }
\NewDocumentCommand \inr		{}		{ \mathsf{inr} }

\DeclareMathOperator \supp 				{ supp }
\NewDocumentCommand	\fsupp	 	{}		{ \mathsf{fs} }
\NewDocumentCommand \perm 		{m}		{ \symfrak{S}_{\fsupp}(#1) }
\NewDocumentCommand \abs 		{m}		{ \langle {#1} \rangle }
\NewDocumentCommand \Abs 		{o}		{ [\Vars] \IfValueT{#1}{#1} }
\NewDocumentCommand \concr 		{}		{ \mathop{@} }
\NewDocumentCommand \disjsupp 	{}		{ \mathrel{\#} }

\NewDocumentCommand \termfunc 		{o}		{%
	\mathcal{F}_{ \IfValueTF{#1}{#1}{\Sigma} }
}
\NewDocumentCommand \lfunc 			{o} 	{ \termfunc[λ\IfValueT{#1}{#1}] }
\NewDocumentCommand \terms			{o}		{%
	\mathcal{T}_{ \IfValueTF{#1}{#1}{\Sigma} }
}
\NewDocumentCommand \termsi			{o}		{%
	\mathcal{T}_{ \IfValueTF{#1}{#1}{\Sigma} }^{\infty}
}
\NewDocumentCommand \qtermfunc 		{o}		{%
	\mathcal{Q}_{ \IfValueTF{#1}{#1}{\Sigma} }
}

\NewDocumentCommand \trunc 			{om}	{
	\left\lfloor #2 \right\rfloor_{ \IfValueTF{#1}{#1}{n} }
}
\NewDocumentCommand \anm 			{o}		{ \symbb{d}\IfValueT{#1}{^{#1}} }

\NewDocumentCommand \substmap 		{}		{ \mathsf{subst} }

\NewDocumentCommand \Vars 		{}		{ \mathcal{V} }
\NewDocumentCommand \lcalc		{} 		{ \Lambda }

\NewDocumentCommand \linf 		{o} 	{
	\Lambda^{ \IfValueTF{#1}{#1}{\infty} }
}
\NewDocumentCommand \lbinf 		{o} 	{
	\Lambda^{ \IfValueTF{#1}{#1}{\infty}_{\bot} }
}

\NewDocumentCommand	\ffv 	{}		{ \mathrm{ffv} }

\DeclareDocumentCommand \subst {mmo} {
	#1 \!\left[ #2 / \IfValueTF{#3}{#3}{x} \right]
}

\NewDocumentCommand \flb 	{o}		{ \longrightarrow_{\beta\IfValueT{#1}{#1}} }
\NewDocumentCommand \flbb 	{o}		{ \flb[\bot\IfValueT{#1}{#1}] }
\NewDocumentCommand \flbs	{o}		{ \flb[#1]^* }
\NewDocumentCommand \flbbs 	{o}		{ \flbs[\bot\IfValueT{#1}{#1}] }

\NewDocumentCommand \flbi	{o}		{ \flb[#1]^{\infty} }
\NewDocumentCommand \flbbi 	{o}		{ \flbi[\bot\IfValueT{#1}{#1}] }

\NewDocumentCommand \chn 	{o}		{ \underline{ \IfValueTF{#1}{#1}{n} } }

\NewDocumentEnvironment{nstack}	{m}		{
	\let\oldarraystretch\arraystretch
	\renewcommand{\arraystretch}{0.7}
	\begin{array}[#1]{@{}c@{}}
}{
	\end{array}
	\let\arraystretch\oldarraystretch
}

\NewDocumentCommand \inset 		{m}		{ \scalebox{0.7}{$\co1 #1$} }
\NewDocumentCommand \innom 		{m}		{ \scalebox{0.7}{$\co2 #1$} }
\NewDocumentCommand \forget 	{m}		{ {\co{2!60} U(} #1 {\co{2!60} )} }

\newcommand{\autocite}[1]{\cite{#1}}

\title{%
	Nominal Algebraic-Coalgebraic Data Types,\\
	with Applications to Infinitary $\bm{\lambda}$-Calculi
	\vspace*{1ex}
	
	\Large A short\thanks{%
		A long version of this abstract can be downloaded from
		\href{https://www.i2m.univ-amu.fr/perso/remy.cerda/}%
		{the author's webpage},
		and appeared as a part of \cite{Cerda24}.
	}
\phantom{\textsuperscript{*}}	fanfiction on \cite{KurzPetrisanAl13}
}
\author{%
	Rémy Cerda\thanks{%
		The author wishes to thank Dimitri Ara, 
		whose categorical knowledge was a great help,
		as well as Léo Hubert, Étienne Miquey and Lionel Vaux Auclair
		for helpful and stimulating discussions,
		and an anonymous referee who suggested several
		highly pertinent references.
	}
	\institute{Aix-Marseille Univ., CNRS, I2M}
	\email{rcerda@math.cnrs.fr}
}

\def\titlerunning{Nominal Algebraic-Coalgebraic Data Types}
\def\authorrunning{R. Cerda}

\hypersetup{allbordercolors=gray!70!blue!30}

\begin{document}

\maketitle

\begin{abstract}
	Ten years ago, it was shown that nominal techniques 
	can be used to design coalgebraic data types with variable binding, 
	so that α-equivalence classes of infinitary terms are directly endowed 
	with a corecursion principle \cite{KurzPetrisanAl13}.
	We introduce \enquote{mixed} binding signatures, 
	as well as the corresponding type of mixed inductive-coinductive terms. 
	We extend the aforementioned work to this setting. 
	In particular, this allows for a nominal description of
	the sets $\linf[abc]$ of $abc$-infinitary λ-terms
	(for $a,b,c \in \{0,1\}$)
	and of capture-avoiding substitution on α-equivalence classes
	of such terms.
\end{abstract}

α-equivalence,
the relation on λ-terms obtained by renaming bound variables,
is central in λ-calculus (as in any syntax with binding):
it is crucially needed in order to define
capture-avoiding substitution in a satisfactory (\ie total) manner,
and thus to define β-reduction.
Even though there are several well-known treatments of it
— \emph{via} the classical \enquote{variable convention} 
\autocite{Barendregt84},
or using \enquote{de Bruijn indices} \autocite{deBruijn72}
more suited to computer-assisted formalisations —
giving abstract and canonical presentations of
the operations of quotient by α-equivalence
and capture-avoiding substitution
has been pursued by several lines of research in the last decades.
Such presentations have been proposed
\emph{via} the introduction of binding algebras \autocite{FioreAl99},
nominal sets \autocite{GabbayPitts02,Pitts13}
or more recently De~Bruijn algebras \autocite{HirschowitzEtAl22}.

In infinitary λ-calculi \autocite{KennawayAl97,Berarducci96},
the precise definition of α-equivalence
is not as standard and straightforward as in a finite setting,
in particular because some issues arise 
from the possibility to encounter terms
containing free occurrences of \emph{all} the available variables.
Applying nominal techniques to the study of infinitary terms
led 
Kurz, Petrişan, Severi, and de Vries
 to establish
a canonical, abstract framework for defining α-equivalence
in a coalgebraic setting \autocite{KurzPetrisanAl12,KurzPetrisanAl13}.

They conclude their work by suggesting that this framework
could be applied not only to the 
\enquote{full} infinitary λ-calculus $\linf[111]$,
but also to its \enquote{mixed} inductive-coinductive variants,
\eg $\linf[001]$ \autocite{KennawayAl97,DalLago16}.
Doing so is the point of this small fanfiction\footnote{%
	By using \href{https://en.wikipedia.org/wiki/Fanfiction}{that word},
	we want to make clear that
	we do not claim much originality in the leading ideas of this work;
	we follow the very same path as \autocite{KurzPetrisanAl13},
	and we perform the necessary adaptions to lift their results
	to an inductive-coinductive setting.
}.
Our contribution is twofold:
\begin{enumerate}
\item We provide an adapted framework
	for general \enquote{mixed} terms with binding
	by introducing \emph{mixed binding signatures} (\mbs).
	The main difference in their categorical treatment
	is that we replace 1-variable polynomial functors
	with 2-variable ones (\ie bifunctors).
\item We show that the proof of \autocite{KurzPetrisanAl13}
	can be easily adapted to this slightly more general setting.
	As an example, we define capture-avoiding substitution on $\linf[001]$
	by mixed recursion and corecursion.
\end{enumerate}

\section{Mixed binding signatures and mixed terms}
\label{sec:terms}

In this section, we introduce \emph{mixed} binding signatures
as well as the finite and infinitary terms arising from
such a signature.
Then we extend to this setting
all the metric and nominal structures
one considers when dealing with ordinary binding signatures,
and we describe a problem similar to what
\autocite{KurzPetrisanAl13} solves in the ordinary setting.

\subsection{Nominal preliminaries}

Let us first recall a few basic definitions and properties
about nominal sets.
We remain quite superficial, since most of the nominal machinery
is hidden in this paper;
we refer to the excellent summary in 
\cite[Sec.~4]{KurzPetrisanAl13},
from which we take all our notations,
and to the standard literature on the subject \autocite{Pitts13}.

Fix a set $\Vars$ of \emph{variables}\footnote{%
	So far, we do not precise the cardinality of $\Vars$.
	In all what follows, $\Vars$ can be countable or uncountable,
	if not specified.
} and denote by $\perm{\Vars}$ the group of the permutations of $\Vars$
that are generated by transpositions $\transp{x}{x'}$,
\ie such that $\setcomp{x \in \Vars}{\sigma(x) \neq x}$ is finite.
A nominal set $(A,\cdot)$ is a set $A$
equipped with a $\perm\Vars$-action $\cdot$
such that each $a \in A$ is \emph{finitely supported},
\ie there exists a least finite set $\supp(a) \subset \Vars$
such that \[\forall \sigma \in \perm\Vars,\ 
(\forall x \in \supp(a),\ \sigma(x) = x)
\Rightarrow \sigma \cdot a = a.\]
Intuitively, variables in $\supp(a)$ are \enquote{free in $a$}.
Nominal sets together with $\perm\Vars$-equivariant maps
form a category $\NOM$.

The key object in all what follows is the \emph{abstraction functor}
$\Abs - : \NOM \to \NOM$ defined as follows.
Fix a nominal set $(A,\cdot)$.
$\Vars \times A$ is equipped with an equivalence relation $\sim$
defined by
\[	(x,a) \sim (x',a') \quad \text{whenever} \quad
	\exists y \notin \supp(a) \cup \supp(a') \cup \{x,x'\},\ 
	\transp xy \cdot a = \transp {x'}y \cdot a'. \]
The intuition behind $\sim$ is that it equates elements of $A$
modulo renaming of free occurrences of a single given variable.
We denote by $\abs xa$ the class of $(x,a)$ in $(\Vars \times A)/\sim$,
and we define a $\perm\Vars$-action on such classes by
$\sigma \cdot \abs xa \defeq \abs{\sigma(x)} (\sigma \cdot a)$.
The functor $\Abs -$ is defined by $\Abs A \defeq (\Vars \times A)/\sim$
on objects, and $\Abs f : \abs xa \mapsto \abs x f(a)$ on morphisms.

The reverse construction is \emph{concretion}, \ie 
the partial equivariant map $\Abs A \times \Vars \to A$
defined by $(\abs xa,y) \mapsto \abs xa \concr y \defeq
\transp xy \cdot a$ for $y \notin \supp(\abs xa)$.
In particular, given such a $y$ we can abstract again on $y$
and form $\abs y \left( \abs xa \concr y \right) = \abs xa$.

\subsection{Categorical preliminaries}

In all what follows and if not specified,
the category $\CC$ is either $\SET$ or $\NOM$.

Given an endofunctor $F : \CC \to \CC$,
an \emph{$F$-algebra} $(A,\alpha)$ is an object $A \in \CC$
together with an arrow $\alpha : FA \to A$.
An algebra morphism $(A,\alpha) \to (B,\beta)$
is an arrow $f : A \to B$
such that $\beta \circ Ff = f \circ \alpha$ in $\CC$.
This defines a category of $F$-algebras.
When this category has an initial object,
it is called the \emph{initial algebra} of $F$
and is denoted by $(\lfp X.FX, \ialg{F})$,
or only $\lfp X.FX$ when there is no ambiguity.
Dualising all these definitions, one obtains
a notion of \emph{terminal coalgebra} for an endofunctor $F$,
denoted by $\gfp X.FX$ when it exists.

Lambek's lemma \autocite{Lambek68}
states that the arrows supporting 
initial algebras and terminal coalgebras
are isomorphisms.
This implies that an initial algebra is a coalgebra,
and that a terminal coalgebra is an algebra.
As a consequence, there is a canonical morphism
$\lfp X.FX \rightarrowtail \gfp X.FX$.

All the functors that we will consider will have a polynomial shape
that makes them \emph{$\omega$-cocontinuous},
\ie they commute to colimits of $\omega$-chains\footnote{%
	What we have in mind is the naive notion of polynomial,
	as considered for instance by
	Adámek, Milius, and Moss \cite{AdamekAl18} 
	 or Métayer \cite{Metayer03}. 
	In particular, the broader notion known as \emph{polynomial functors}
	encompasses functors with infinite powers,
	which prevents $\omega$-cocontinuity in general.
	See Kock \cite[§~1.7.3]{Kock09} for a discussion.
}.
This entails the existence of their initial algebra.
Given an $\omega$-cocontinuous bifunctor $F : \CC \times \CC \to \CC$,
one can take the initial algebra in the first variable:
this gives rise to an $\omega$-cocontinuous functor
$\lfp X.F(X,-) : \CC \to \CC$.

\begin{lem}[diagonal identity] \label{lem:diagonal}
	Given an $\omega$-cocontinuous functor $F : \CC \times \CC \to \CC$,
	\[\lfp Y.\lfp X.F(X,Y) = \lfp Z.F(Z,Z)\]
	in the category of $F\diagfunc$-algebras,
	where $\diagfunc : X \mapsto (X,X)$ is the diagonal functor.
\end{lem}
This lemma is standard, and has been first proved
in a categorical setting by  Lehmann and Smyth \cite{LehmannSmyth81}. 
An alternative proof is proposed in the appendices
of the long version of this abstract.

\subsection{\mbs and raw terms}

Binding signatures \autocite{Plotkin90,FioreAl99}
provide a general description of term (co)algebras
with binding operators.
Let us quickly recall their main properties.
A \emph{binding signature} (\bs) is a couple $(\Sigma, \arite)$
where $\Sigma$ is a set at most countable of constructors,
and $\arite : \Sigma \to \Nat^*$ is a function
indicating the binding arity of each argument of each constructor.
Given a \bs $(\Sigma, \arite)$,
its term functor $\termfunc : \CC \to \CC$
is defined by
\[	\termfunc X \defeq \Vars + \coprod_{\substack{
		\opcons \in \Sigma \\ \arite(\opcons) = (n_1,\dots,n_k)
	}} \prod_{i=1}^{k} \Vars^{n_i} \times X. \]
The sets of raw (\ie not quotiented by α-equivalence)
finite and infinitary
terms on $(\Sigma, \arite)$ are then defined by
$\terms \defeq \lfp X.\termfunc X$ and
$\termsi \defeq \gfp X.\termfunc X$ (in $\SET$).
Notice that these (co)algebras always exist,
thanks to the polynomial shape of $\termfunc$.
A typical example is the signature:
	$\Sigma_λ \defeq \{ λ,@ \}$ with
	$\arite(λ) \defeq (1)$ and
	$\arite(@) \defeq (0,0)$,
such that $\terms[λ]$ is the algebra $\lcalc$ of all finite λ-terms,
and $\termsi[λ]$ is the coalgebra $\linf[111]$
of all (full) infinitary λ-terms.

We want to tweak this definition
in order to be able to design
mixed inductive-coinductive data types with binding\footnote{%
	Existing generalisations of binding signatures
	go way beyond our modest extension,
	that could certainly be reformulated in a broader setting ---
	see \eg Power  
	\cite{Power07} (whose work subsumes both
	Fiore, Plotkin, and Turi's
	 binding algebras
	and 
	Gabbay and Pitts'
	 nominal sets), 
	as well as Adámek, Milius, and Velebil 
	\cite{AdamekMiVe09} or 
	Arkor 
	\cite{Arkor22}.
	However, it is not completely clear to us
	whether these abstract frameworks encompass
	coinductive syntax in the way we want to construct it,
	without any further work.
}.
An elementary example of such a mixing (with no binding)
is the type of \emph{right-infinitary} binary trees:
the set of all infinitary binary trees such that
each infinite branch contains infinitely many right edges.
This type can be defined as $\gfp Y.\lfp X.1+X \times Y$ in $\SET$.
Our aim is to be able to express such a construction
when some constructors bind variables
(and then investigate the quotient by α-equivalence).

\begin{defi}[mixed binding signature]
	A \emph{mixed binding signature} (\mbs) is a couple $(\Sigma, \arite)$
	where $\Sigma$ is a set at most countable of constructors,
	and $\arite : \Sigma \to (\Nat \times \Bool)^*$ is an arity function.
\end{defi}

$\Bool$ denotes the set of booleans:
each argument of each constructor is marked with
a boolean describing its (co)inductive behaviour.
This intuition is driving the following definitions,
that allow to define \emph{mixed} terms on a \mbs.

\begin{defi}[term functor of a \mbs] \label{def:termfunc}
	The polynomial \emph{term functor} associated to $(\Sigma, \arite)$
	is the $\CC$-bifunctor $\termfunc$ defined by:
	\[	\termfunc (X,Y) \defeq \Vars +
		\coprod_{ \substack{	\opcons \in \Sigma \\ 
								\arite(\opcons) = ((n_1,b_1),\dots,(n_k,b_k))
		}} \prod_{i=1}^{k} \Vars^{n_i} \times \pi_{b_i}(X,Y) \]
	where $\pi_0$ and $\pi_1$ are the projections.
\end{defi}

\Cref{lem:diagonal} ensures that there is a unique
notion of \enquote{fully initial} algebra
on a bifunctor,
hence the definition of raw terms on a \mbs.

\begin{defi}[raw terms on a \mbs] \label{def:raw_terms}
	The sets $\terms$ of \emph{raw finite terms}
	and $\termsi$ of \emph{raw mixed terms}
	on $(\Sigma, \arite)$ are defined by:
	\[	\terms \defeq \lfp Z.\termfunc (Z,Z) \qquad 
		\termsi \defeq \gfp Y.\lfp X.\termfunc (X,Y).	\]
\end{defi}

\begin{figure} \centering 
	\begin{prooftree}[center=false]
		\hypo{ {\co1 x} \in \Vars }
		\infer1{ {\co1 x} \in \termsi }
	\end{prooftree}
	\qquad 
	\begin{prooftree}[center=false]
		\hypo{ {\co1 t} \in \termsi }
		\infer1{ \mathrel{{\co2 \later_0}} {\co1 t} \in \termsi }
	\end{prooftree}
	\qquad
	\begin{prooftree}[center=false]
		\hypo{ {\co1 t} \in \termsi }
		\infer[double]1{ \mathrel{{\co2 \later_1}} {\co1 t} \in \termsi }
	\end{prooftree}
	\vspace{2ex}
	
	\begin{prooftree}[center=false]
		\hypo{ {\co1 \bar{x_1}} \in \Vars^{n_1} }
		\hypo{ \cdots }
		\hypo{ {\co1 \bar{x_k}} \in \Vars^{n_k} }
		\hypo{ \mathrel{{\co2 \later_{b_1}}} {\co1 t_1} \in \termsi }
		\hypo{ \cdots }
		\hypo{ \mathrel{{\co2 \later_{b_k}}} {\co1 t_k} \in \termsi }
		\infer6{
			\opcons\left(\sabs{{\co1\bar{x_1}}} {{\co1 t_1}},
						\dots, 
						\sabs{{\co1\bar{x_k}}} {{\co1 t_k}}
			\right) \in \termsi }
	\end{prooftree}
	
	{\small for each $\opcons \in \Sigma$, having
	$\arite(\opcons) = ((n_1,{\co2 b_1}),\dots,(n_k,{\co2 b_k}))$}
	\caption{Formal system defining $\termsi$
		for a \mbs $(\Sigma, \arite)$.
		The simple rules are inductive, the double one is coinductive;
		for similar systems, see \autocite{DalLago16,CerdaVaux23}.}
	\label{fig:rulesMixedRawTerms}
\end{figure}

\begin{nota}
	We can describe $\termsi$
	by means of a (mixed) formal system of derivation rules,
	as proposed in \cref{fig:rulesMixedRawTerms}.
	We use the symbols $\later_0$ and $\later_1$ to distinguish
	between the inductive and coinductive calls.
	$\later_1$ is usually called the \emph{later} modality
	\autocite{Nakano00,AppelAl07};
	a derivation of $\later_1 P$
	is a derivation of $P$ under an additional coinductive guard.
	The modality $\later_0$ could be omitted,
	but we write it to keep the notations symmetric.
\end{nota}

\begin{figure}
\begin{center}
	\hspace*{\stretch{2}}
	\begin{prooftree}[center=false]
		\hypo{ {\co1 x} \in \Vars }
		\infer1{ {\co1 x} \in \linf[001] }
	\end{prooftree}
	\hspace*{\stretch{1}}
	\begin{prooftree}[center=false]
		\hypo{ {\co1 M} \in \linf[001] }
		\infer[double]1{ \mathrel{{\co2 \later}} {\co1 M} \in \linf[001] }
	\end{prooftree}
	\hspace*{\stretch{1}}
	\begin{prooftree}[center=false]
		\hypo{ {\co1 x} \in \Vars }
		\hypo{ {\co1 M} \in \linf[001] }
		\infer2{ λ( \sabs{{\co1 x}} {{\co1 M}} ) \in \linf[001] }
	\end{prooftree}
	\hspace*{\stretch{1}}
	\begin{prooftree}[center=false]
		\hypo{ {\co1 M} \in \linf[001] }
		\hypo{ \mathrel{{\co2 \later}} {\co1 N} \in \linf[001] }
		\infer2{ @ ({{\co1 M}}, {{\co1 N}} ) \in \linf[001] }
	\end{prooftree}
	\hspace*{\stretch{2}}
	
	\caption{A simplified mixed formal system defining $\linf[001]$.}
	\label{fig:rulesMixedRawLamTerms}
\end{center}
\end{figure}

\begin{exa}[mixed infinitary λ-terms] \label{exa:abcSignatures}
	For $a,b,c \in \Bool$, the \mbs $(\Sigma_λ,\arite_{abc})$ is defined by:
	\[	\Sigma_λ \defeq \{ λ, @ \} \qquad 
		\arite_{abc}(λ) \defeq ((1,a)) \qquad
		\arite_{abc}(@) \defeq ((0,b), (0,c)). \]
	For any $a,b,c$, $\terms[λabc]$
	is the algebra $\lcalc$ of finite λ-terms
	and $\termsi[λabc]$ is the coalgebra of $abc$-infinitary λ-terms.
	For instance, the 001-infinitary λ-terms
	are described by \cref{fig:rulesMixedRawLamTerms}.
\end{exa}

\subsection{Metric completion}

Take $\CC$ to be $\SET$. Following a standard path,
we define the Arnold-Nivat metric \autocite{ArnoldNivat80}
on both $\terms$ and $\termsi$.
To do so, we use the following notion of truncation,
adapted to a mixed inductive-coinductive setting.

\begin{defi}[truncation]
	Given an integer $n \in \Nat$ and a term $t$
	in either $\terms$ or $\termsi$,
	the \emph{mixed truncation} at depth $n$ of $t$ is the object\footnote{%
		We try not to be too formal here.
		In the following we manipulate truncations as if they were
		finite terms on $\Sigma \cup \{\ast\}$,
		where $\ast$ is a new constant.
	}
	$\trunc t \in (\lfp X.\termfunc(X,-))^n 1$ defined by induction by:
	\begin{align*}
		\trunc[\co2 0]   {{\co0 t}} 	& \defeq * \\
		\trunc[\co2 n+1] {{\co1 x}} 	& \defeq {\co1 x} \\
		\trunc[\co2 n+1] {{\co0
			\opcons\left( \sabs{\bar{x_1}} {\co1 t_1}, \dots, 
			\sabs{\bar{x_k}} {\co1 t_k} \right)
			}}							& \defeq
			\opcons \left( \sabs{\bar{x_1}}
				\trunc[\co2 n+1-b_1]{{\co1 t_1}}, \dots, 
			\sabs{\bar{x_k}} \trunc[\co2 n+1-b_k]{{\co1 t_k}} \right)
	\end{align*}
	where ${\co2 b_i} = \pi_1 \arite(\opcons)_i$.
\end{defi}

Notice that the definition is by double induction,
on $n$ and on $t$ (even if the latter is taken in $\termsi$):
in the inductive arguments of $\opcons$ we proceed by induction on $t$,
in its coinductive arguments we proceed by induction on $n$.

\begin{defi}[Arnold-Nivat metric] \label{def:anMetric}
	The \emph{Arnold-Nivat metric} on $\terms$ and $\termsi$
	is the mapping
	$\anm : \termsi \times \termsi \to \Reals[+]$ defined by
	$\anm(t,u) \defeq
		\inf\setcomp{ 2^{-n} }{ n\in\Nat,\ \trunc t = \trunc u }$.
\end{defi}

The unique notation is unambiguous,
since the canonical inclusion $\terms \rightarrowtail \termsi$
preserves the truncations.
The following fact is a translation of 
\cite[Th.~3.2]{Barr93},
using \cref{lem:diagonal}.
It expresses the equivalence of our coinductive definition of $\termsi$
and the historical topological point of view
\autocite{KennawayAl97}.

\begin{lem} \label{lem:completionSetBarr}
	$\termsi$ is the Cauchy completion of $\terms$
	with respect to $\anm$.
	Furthermore, the completion is carried by the canonical arrow
	$\terms \rightarrowtail \termsi$.
\end{lem}

\begin{exa}
	The eight Arnold-Nivat metrics $\anm[abc]$
	corresponding to the signatures from \cref{exa:abcSignatures}
	are exactly those considered in the original definition
	of infinitary λ-calculi \autocite{KennawayAl97}.
	Hence our coinductive definition of $\linf[abc]$
	coincides with the historical, topological definition.
\end{exa}

\subsection{α-equivalence}

α-equivalence is the equivalence relation generated
on some term (co)algebra by renaming all bound variables.
Let us recall how this can be reformulated in a nominal setting
(for finite terms):
given a \bs or a \mbs $(\Sigma, \arite)$,
the finite term algebra $\terms$ can be endowed with
a $\perm\Vars$-action $\cdot$ inductively defined by:
\begin{equation} \label{eq:sigmavaction}
\begin{array}{rcl}
	\sigma \cdot {\co1 x} & \defeq & \sigma({\co1 x}) \\
	\sigma \cdot \opcons\left( \sabs{{\co2\bar{x_1}}} {\co1 t_1}, \dots\right)
		& \defeq & \opcons\left( \sabs{\sigma({\co2 \bar{x_1}})}
		\sigma \cdot {\co1 t_1}, \dots \right),
\end{array}
\end{equation}
where permutations act pointwise on the sequences $\bar{x_i}$.
This defines a nominal set $(\terms,\cdot)$.
The α-equivalence relation is then defined by:
\[
	\begin{prooftree}[center=false]
		\infer0{ {\co1 x} \aeq {\co1 x} \vphantom{\sabs{\bar{z_i}}} }
	\end{prooftree}
	\qquad 
	\begin{prooftree}[center=false]
		\hypo{ \left(
			\transp{{\co2\bar{x_i}}}{{\co2\bar{z_i}}} \cdot {\co1 t_i}
			\aeq
			\transp{{\co2\bar{y_i}}}{{\co2\bar{z_i}}} \cdot {\co1 u_i}
			\text{ for fresh $\co2\bar{z_i}$}
			\right)_{i=1}^k }
		\infer1{
			\opcons\left( \sabs{{\co2\bar{x_1}}}{{\co1 t_1}},\dots \right)
			\aeq 
			\opcons\left( \sabs{{\co2\bar{y_1}}}{{\co1 u_1}},\dots \right)
		}
	\end{prooftree}
\]
where the permutation $\transp{\bar{x_i}}{\bar{z_i}}$ is the composition
of the transpositions $\transp{x_i}{z_i}$.
This equivalence relation is compatible with $\cdot$,
thus there is an induced nominal structure $(\terms\qbalpha, \cdot)$.

An important theorem by Gabbay and Pitts
\cite[Th.~8.15]{GabbayPitts02,Pitts13}
can be straightforwardly transported to our mixed setting.

\begin{defi}[quotient term functor of a \mbs]
	The polynomial \emph{quotient term functor}
	associated to $(\Sigma, \arite)$
	is the $\NOM$-bifunctor $\qtermfunc$ defined by:
	\[	\qtermfunc (X,Y) \defeq \Vars +
		\coprod_{ \substack{	\opcons \in \Sigma \\ 
								\arite(\opcons) = ((n_1,b_1),\dots,(n_k,b_k))
		}} \prod_{i=1}^{k} \Abs^{n_i} \pi_{b_i}(X,Y). \]
\end{defi}

\begin{thm}[nominal algebraic types on a \mbs] \label{thm:nomAlgTypes}
	Given a \mbs $(\Sigma, \arite)$, then
	$\terms = \lfp Z.\termfunc(Z,Z)$ and
	$\terms\qbalpha = \lfp Z.\qtermfunc(Z,Z)$
	in $\NOM$.
\end{thm}

The first identity might seem tautologic
because of the overloaded the notation $\termfunc$;
if we distinguish between $\termfunc^{\SET}$ and $\termfunc^{\NOM}$
it becomes
$(\lfp Z.\termfunc^{\SET}(Z,Z), \cdot) = \lfp Z.\termfunc^{\NOM}(Z,Z)$,
where the former nominal structure was built
in \cref{eq:sigmavaction}.

\subsection{Towards commutation (or not)}

For now, we have built the following diagram (in $\SET$):
\begin{equation} \label[diagram]{thediagram1}
\begin{tikzcd}[column sep=15mm, row sep=5mm]
	\begin{nstack}{b}
		\innom{\forget{ \lfp Z.\termfunc(Z,Z) }} \\
		\inset{ \lfp Z.\termfunc(Z,Z) } \\
		\terms
	\end{nstack}
		\arrow[tail]{r}{\co1 \text{compl.}}
		\arrow[two heads]{d}
&	\begin{nstack}{b}
		\inset{ \gfp Y.\lfp X.\termfunc(X,Y) } \\
		\termsi
	\end{nstack}
\\	\begin{nstack}{t}
		\terms\qbalpha \\
		\innom{\forget{ \lfp Z.\qtermfunc(Z,Z) }}
	\end{nstack}
\end{tikzcd}
\end{equation}
The sets are annotated with their descriptions as (co)algebras
{\co1 in $\SET$} and {\co2 in $\NOM$}
($U$ is the forgetful functor $\NOM \to \SET$).
The horizontal arrow is the metric completion
given by \cref{lem:completionSetBarr},
the vertical surjection is the quotient by α-equivalence
given by \cref{thm:nomAlgTypes}.
Our goal is to close the square with an object
containing α-equivalence classes of mixed terms;
we hope to obtain a nominal presentation of this object.
To do so, we keep adapting the definitions of
\cite{KurzPetrisanAl13} to our mixed setting:
\begin{itemize}
\item $\termsi$ can be equipped with a $\perm\Vars$-action
	in the same way as we did in \cref{eq:sigmavaction}
	for the finitary setting,
	by just making the definition coinductive;
	however, this does not define a nominal set any more
	since some infinitary terms are not finitely supported
	(the support of a term being the set of the variables
	occurring in it).
\item As a consequence, we cannot directly use a nominal set structure
	to extend the definition of α-equivalence to $\termsi$.
	Instead, we lift the α-equivalence of $\terms$
	by using the truncations:
	two mixed terms $t,u \in \termsi$ are then said to be α-equivalent
	if $\forall n\in\Nat,\ \trunc t \aeq \trunc u$.
\item We also define a metric on $\terms\qbalpha$
	as in \cref{def:anMetric}:
	$\anm_α(t,u) \defeq
		\inf\setcomp{ 2^{-n} }{ n\in\Nat,\ \trunc t \aeq \trunc u }$.
	Then $(\terms\qbalpha)^\infty$ is the metric completion
	of $\terms\qbalpha$ with respect to $\anm_α$.
\end{itemize}
These constructions extend \cref{thediagram1} as follows:
\begin{equation} \label[diagram]{thediagram2}
\begin{tikzcd}[column sep=15mm, row sep=5mm]
	\begin{nstack}{b}
		\innom{\forget{ \lfp Z.\termfunc(Z,Z) }} \\
		\inset{ \lfp Z.\termfunc(Z,Z) } \\
		\terms
	\end{nstack}
		\arrow[tail]{r}{\co1 \text{compl.}}
		\arrow[two heads]{dd}
&	\begin{nstack}{b}
		\inset{ \gfp Y.\lfp X.\termfunc(X,Y) } \\
		\termsi
	\end{nstack}
		\arrow[two heads]{d}
\\&	\termsi\qbalpha
		\arrow[hook]{d}{\text{\bf ?}}
\\	\begin{nstack}{t}
		\terms\qbalpha \\
		\innom{\forget{ \lfp Z.\qtermfunc(Z,Z) }}
	\end{nstack}
		\arrow[tail]{r}{\co1 \text{compl.}}
&	(\terms\qbalpha)^\infty
\end{tikzcd}
\end{equation}
The existence of an inclusion
$\stackrel{\text{\bf ?}}{\hookrightarrow}$
is straightforward,
but we would like an isomorphism instead.
Unfortunately, it is not the case in general,
unless the signature is trivial in the following meaning.

\begin{defi}[non-trivial \mbs] \label{def:nontrivial}
	A \mbs $(\Sigma,\arite)$ is \emph{non-trivial}
	if there are constructors 
	$\oplam$, $\opnode$, $\opdig \in \Sigma$ such that:
	\begin{enumerate}
	\item $\oplam$ has a binding argument,
		\ie $\pi_0(\arite(\oplam)_i) \geq 1$ for some index $i$;
	\item $\opnode$ has at least two arguments,
		\ie $\arite(\opnode)$ is of length greater than~2; 
	\item $\opdig$ has a coinductive argument,
		\ie $\pi_1(\arite(\opdig)_i) = 1$ for some index $i$.
	\end{enumerate}
\end{defi}

If the signature is trivial, it does not make sense to consider
all the machinery defined here:
if there is no binder then $\aeq$ amounts to equality,
if there are only unary and constant constructors
then there is at most one variable in each term,
and if there is no coinductive constructor then the metric is discrete.
In all three cases, $(\termsi\qbalpha) \cong (\terms\qbalpha)^\infty$
for degenerate reasons.
Otherwise, the cardinality of $\Vars$ is determining,
as \cref{thm:uncountableCountable} shows.

\begin{thm} \label{thm:uncountableCountable}
	Let $(\Sigma,\arite)$ be a non-trivial \mbs. Then
	$(\termsi\qbalpha) \cong (\terms\qbalpha)^\infty$
	iff $\Vars$ is uncountable.
\end{thm}

Our goal is not really fulfilled:
we have a commutative square only if $\Vars$ is uncountable,
which is not satisfactory in practice
since implementation concerns suggest to
consider countably many variables.
In addition, none of the sets involved can be endowed
with a reasonable nominal structure.

\section{The nominal coalgebra of α-equivalence classes of mixed terms}

In this second part,
we show that 
Kurz, Petrişan, Severi, and de Vries'
 theorem
has a mixed counterpart.
Then we use this result to define substitution
on mixed terms by nested recursion and corecursion.

\subsection{Nominal mixed types}

The following structure is, once again,
extended to the setting of mixed terms:
\begin{itemize}
\item Given a set $S$ equipped with a $\perm\Vars$-action,
	$S_{\fsupp}$ is the subset of finitely supported elements of $S$.
	It carries a nominal set structure.
	In particular $(\termsi)_{\fsupp}$ is the nominal set of
	the finitely supported raw terms in $\termsi$,
	and $(\terms\qbalpha)^{\infty}_{\fsupp}$ is the nominal set of
	finitely supported α-equivalence classes in $(\terms\qbalpha)^{\infty}$.
\item $(\termsi)_\ffv$ denotes the set of infinitary terms
	having finitely many free variables.
\end{itemize}

Recall also that given a nominal metric space
(\ie a nominal space equipped with an equivariant metric),
its nominal metric completion is built by adding
the limits of all finitely supported Cauchy sequences
(\ie sequences of terms such that their supports are all
contained in a common finite set).

Let us state the main theorem of our fanfiction without delay,
as well as its crucial corollary.

\begin{thm}[nominal mixed terms on a \mbs] \label{thm:nomMixedTypes}
	Let \mbs $(\Sigma,\arite)$ be a \mbs. Then:
	\begin{enumerate}
	\item The nominal set $(\termsi)_{\fsupp}$ is
		the nominal metric completion of $\terms$,
		as well as the terminal coalgebra
		$\gfp Y.\lfp X.\termfunc(X,Y)$.
	\item Similarly, the nominal set $(\terms\qbalpha)^{\infty}_{\fsupp}$ is
		the nominal metric completion of $\terms\qbalpha$,
		as well as the terminal coalgebra
		$\gfp Y.\lfp X.\qtermfunc(X,Y)$.
	\item The following diagram commutes in $\SET$:
	\[\begin{tikzcd}[column sep=12mm]
		\begin{nstack}{b}
		\innom{\forget{ \lfp Z.\termfunc(Z,Z) }} \\
		\inset{ \lfp Z.\termfunc(Z,Z) } \\
		\terms
		\end{nstack}
			\arrow{r}{\co2 \text{nom.}}[below]{\co2 \text{compl.}}
			\arrow[two heads]{d}
			\arrow[tail,bend left=22]{rrr}[description]{\co1 \text{compl.}}
	&	\begin{nstack}{b}
		\innom{\forget{ \gfp Y.\lfp X.\termfunc(X,Y) }} \\
		(\termsi)_\fsupp
		\end{nstack}
			\arrow[hook]{r}
			\arrow{d}
	&	(\termsi)_\ffv
			\arrow[hook]{r}
			\arrow[two heads]{d}
	&	\begin{nstack}{b}
		\inset{ \gfp Y.\lfp X.\termfunc(X,Y) } \\
		\termsi
		\end{nstack}
			\arrow{d}
	\\	\begin{nstack}{t}
		\terms\qbalpha \\
		\innom{\forget{ \lfp Z.\qtermfunc(Z,Z) }}
		\end{nstack}
			\arrow{r}{\co2 \text{nom.}}[below]{\co2 \text{compl.}}
			\arrow[tail,bend right=22]{rrr}[description]{\co1 \text{compl.}}
	&	\begin{nstack}{t}
		(\terms\qbalpha)^\infty_\fsupp \\
		\innom{\forget{ \gfp Y.\lfp X.\qtermfunc(X,Y) }}
		\end{nstack}
			\arrow[equal]{r}
	&	(\termsi)_\ffv\qbalpha
			\arrow[hook]{r}
	&	(\terms\qbalpha)^\infty
			\arrow[phantom]{ul}[very near end]{\lrcorner}
	\end{tikzcd}\]
	\end{enumerate}
\end{thm}

\begin{cor} \label{cor:nomMixedTypes}
	The nominal set $(\termsi)_\ffv\qbalpha$ is the terminal coalgebra
	$\gfp Y.\lfp X.\qtermfunc(X,Y)$.
\end{cor}

These results are direct counterparts to
Remark~5.30, Theorem~5.34 and Corollary~5.35
from \autocite{KurzPetrisanAl13},
and the diagram we provide is exactly the same
as their diagram~5.20.
The only difference here is that we take the terminal coalgebra
of $\lfp X.\termfunc(X,-)$ and $\lfp X.\qtermfunc(X,-)$,
instead of $\termfunc$ and $\qtermfunc$ themselves.
What we need to show is that all the technical developments
of \autocite{KurzPetrisanAl13} remain applicable\footnote{%
	During the writing of this paper,
	we came up with an explicit construction of
	our mixed terms as purely coinductive terms
	on a modified binding signature.
	From this, one gets an alternative proof of the theorem.
	Even if it is not useful for our purposes,
	we provide this construction in the appendices
	of the long version of this abstract,
	just in case.
}.

\begin{lem} \label{lem:polynomes}
	Let $F : \NOM \times \NOM \to \NOM$ be polynomial
	in the following sense:
	there are a countable set $I$ and families
	$\setcomp{ k_i \in \Nat }{ i \in I }$,
	$\setcomp{ m_{ij} \in \Nat }{
		\begin{smallmatrix*}[l] i\in I \\ 1\leq j \leq k_i \end{smallmatrix*}
	}$ and
	$\setcomp{ b_{ij} \in \Bool }{ 
		\begin{smallmatrix*}[l] i\in I \\ 1\leq j \leq k_i \end{smallmatrix*}
	}$ such that
	\[	F = K + \coprod_{i \in I} \prod_{j=1}^{k_i} 
		M^{m_{ij}} \pi_{b_{ij}} \]
	where $\pi_0$ and $\pi_1$ denote the projections,
	$M : \NOM \to \NOM$ is a fixed functor commuting to directed colimits,
	and $K$ is a fixed constant functor.
	Then $\lfp X.F(X,-)$ exists
	and can be obtained from the following grammar
	(up to isomorphism):
	\begin{equation}
		\begin{inductivedef}
		G & \id \sep K \sep MG \sep \coprod G \sep G \times G
		\end{inductivedef}
		\tag{$\Gamma_1$} \label[grammar]{grammar:1}
	\end{equation}
	where $\coprod$ denotes at most countable coproducts.
\end{lem}

Using the lemma,
the proof of \cref{thm:nomMixedTypes,cor:nomMixedTypes}
is straightforward:
taking $K$ to be the constant functor $\Vars$,
and $M$ to be either $\Vars\times-$ or $\Abs$,
we just showed that $\lfp X.\termfunc(X,-)$
and $\lfp X.\qtermfunc(X,-)$
fulfill the requirements of 
\cite[Prop.~5.6]{KurzPetrisanAl13}.

\begin{exa}
	The nominal set $\linf[001]_\ffv\qbalpha$
	of α-equivalence classes of 001-infinitary λ-terms
	having finitely many free variables
	is the terminal coalgebra $\gfp Y.\lfp X.\Vars + \Abs X + X\times Y$.
\end{exa}

\subsection{Capture-avoiding substitution for mixed types}

We fix a \mbs $(\Sigma,\arite)$, 
and we write $\termsi[α]$ for $\gfp Y.\lfp X.\qtermfunc(X,Y)$.
We want to define capture-avoiding substitution as a map
$\substmap : \termsi[α] \times \Vars \times \termsi[α] \to \termsi[α]$
in $\NOM$.

As in 
\cite[Def.~6.2]{KurzPetrisanAl13},
we shall use the corecursion principle of 
\cite[Lem.~2.1]{Moss01}.
However, this is not enough any more:
we also have to scan the inductive structure 
separating two coinductive constructors and, 
since this structure may contain variables
(in fact \emph{all} the variables appear
in these \enquote{inductive layers}),
perform substitution recursively on it too.

\begin{nota}
	When we consider a coproduct $A+B$, we write $\inl$ and $\inr$
	for the left and right injections.
	Similarly, we write $\invar$ and $\incons$ the injections
	in initial algebras of the form $\lfp X.\qtermfunc(X,Y)$.
	We omit the composition by $\fold$ for the sake of readability.
\end{nota}

\begin{nota}
	It is easy to show that $\NOM$-endofunctors obtained
	from (\cref{grammar:1}) are strong, hence we denote:
	\begin{itemize}
	\item by $\tau_{A,B} : \Abs A \times B \to \Abs (A \times B)$
		the strength defined by
		$(\abs x a,b) \mapsto \abs z (\abs xa \concr z, b)$,
	\item by $\tau$ the strength
		$\tau_{\termsi[α], \Vars \times \termsi[α]}$
		and by $\tau_n : \Abs^n \termsi[α] \times \Vars \times \termsi[α]
		\to \Abs^n (\termsi[α] \times \Vars \times \termsi[α])$
		its iteration,
	\item by $\bar{\tau}$ the strength generated for
		$\lfp X.\qtermfunc(X, \termsi[α] + -)$.
	\end{itemize}
\end{nota}
Using these notations, we are finally able to define
capture-avoiding substitution.

\begin{defi}[capture-avoiding substitution] \label{defi:subst}
	\emph{Capture-avoiding substitution} is the map $\substmap$
	defined by:
	\[\begin{tikzcd}[column sep=3cm]
		\termsi[α] \times \Vars \times \termsi[α]
			\arrow[dashed]{r}{\substmap}
			\arrow{d}[left]{\unfold \times\Vars\times\termsi[α]}
	&	\termsi[α]
			\arrow{dd}{\unfold}
	\\	\lfp X.\qtermfunc(X, \termsi[α]) \times \Vars \times \termsi[α]
			\arrow{d}[left]{h'}
	\\	\lfp X.\qtermfunc(X, \termsi[α] + \termsi[α]\times\Vars\times\termsi[α])
			\arrow{r}{\lfp X.\qtermfunc(X, \id + \substmap)}
	&	\lfp X.\qtermfunc(X, \termsi[α])
	\end{tikzcd}\]
	where $h'$ is recursively defined by:
	\begin{align*}
		(\invar(x), x, u) & \mapsto
			\lfp X.\qtermfunc(X,\inl) (\unfold(u)) \\
		(\invar(y), x, u) & \mapsto \invar(y) &
			\text{for $y\neq x$} \\
		\left( \incons\left( \begin{array}{@{}c@{}}
				\vdots \\
				\abs{y_{i,1}} \dots \abs{y_{i,n_i}} t_i, \\
				\vdots \\
				\abs{y_{j,1}} \dots \abs{y_{j,n_j}} t_j \\
				\vdots 
			\end{array} \right), x, u \right)
			& \mapsto \lfp X.\qtermfunc(X,\inr) \left( 
			\begin{array}{@{}c@{}}
				\vdots \\
				\abs{y_{i,1}} \dots \abs{y_{i,n_i}} h'(t_i,x,u), \\
				\vdots \\
				\tau_{n_j}(\abs{y_{j,1}} \dots \abs{y_{j,n_j}} t_j,x,u) \\
				\vdots 
			\end{array} \right)
	\end{align*}
	where $i$ (resp. $j$) represents any index
	such that $b_i = 0$ (resp. $b_j = 1$),
	\ie any inductive (resp. coinductive) position of $\opcons$),
	and where the representatives are taken so that
	$\forall k \in [0,n_i],\ 
	y_{i,k} \disjsupp x \text{ and } y_{i,k} \disjsupp u$.
\end{defi}
In fact the validity of the recursive definition of $h'$
is not immediate; in particular, it is not straightforwardly implied
by Pitts' recursion theorem for nominal algebras 
\cite[Thm.~5.1]{Pitts06}
(see also 
\cite[§~8.5]{Pitts13} for lighter presentation).
This is due to the fact that $h'$ is not purely inductive,
it also inserts $\tau_{n_j}$'s in coinductive positions
(which amounts to modifying the constructors of the local induction step).
This is why a rigorous definition of $h'$ relies on
the following decomposition into a purely inductive $h$,
followed by some work on the coinductive structure of the terms:
\[ h' \defeq \begin{tikzcd}
	\lfp X.\qtermfunc(X, \termsi[α]) \times \Vars \times \termsi[α]
			\arrow{d}[left]{
				\lfp X.\qtermfunc(X,\inr) \times \Vars \times \termsi[α] }
	\\	\lfp X.\qtermfunc(X, \termsi[α] + \termsi[α])
		\times \Vars \times \termsi[α]
			\arrow{d}[left]{ h  \times \Vars \times \termsi[α] }
	\\	\lfp X.\qtermfunc(X, \termsi[α] + \termsi[α])
		\times \Vars \times \termsi[α]
			\arrow{d}[left]{\bar\tau}
	\\	\lfp X.\qtermfunc(X, \termsi[α] + 
		\termsi[α]\times\Vars\times\termsi[α])
\end{tikzcd} \]
where $h : (t,x,u) \mapsto h_{x,u}(t)$ is uniquely defined by recursion by
\begin{align*}
	\invar(x) & \mapsto
		\lfp X.\qtermfunc(X,\inl) (\unfold(u)) \\
	\invar(y) & \mapsto \invar(y) &	\text{for $y\neq x$} \\
	\incons\left( \begin{array}{@{}c@{}}
			\vdots \\
			\abs{y_{i,1}} \dots \abs{y_{i,n_i}} {\co1 t_i}, \\
			\vdots \\
			\abs{y_{j,1}} \dots \abs{y_{j,n_j}} T_j \\
			\vdots 
		\end{array} \right)
		& \mapsto \incons\left( \begin{array}{@{}c@{}}
			\vdots \\
			\abs{y_{i,1}} \dots \abs{y_{i,n_i}} {\co1 h_{x,u}(t_i)}, \\
			\vdots \\
			\abs{y_{j,1}} \dots \abs{y_{j,n_j}} T_j \\
			\vdots 
		\end{array} \right)
\end{align*}
under the hypotheses and notations of \cref{defi:subst},
that ensure that the \enquote{freshness condition for binders}
of Pitt's recursion theorem is satisfied,
hence the well-definedness of $h$.

\begin{exa}
	Let us describe what $h'$ looks like when $\termsi[α]$
	is $\linf[001]_\ffv\qbalpha$:
	\begin{align*}
		(x, x, N) & \mapsto
			\lfp X.\qtermfunc[λ001](X,\inl) (\unfold(N)) \\
		(y, x, N) & \mapsto y & \text{for $y\neq x$} \\
		(λ(y.M), x, N) & \mapsto \lfp X.\qtermfunc[λ001](X,\inr)
			(λ(y.h(M,x,N))) & \text{for $y\neq x$ and $y \notin \fv(N)$} \\
		(@(M_0,M_1), x, N) & \mapsto \lfp X.\qtermfunc[λ001](X,\inr)
			\left( @\left( h(M_0,x,N), (M_1,x,N) \right) \right),
	\end{align*}
	where we omitted the injections.
	Finally we obtain the expected recursive-corecursive
	definition of capture-avoiding substitution:
	\begin{align*}
		\substmap(x,x,N) & \defeq N \\
		\substmap(y,x,N) & \defeq y & \text{for $y \neq x$} \\
		\substmap(λ(y.M),x,N) & \defeq λ(y.\substmap(M,x,N)) &
			\text{for $y\neq x$ and $y \notin \fv(N)$} \\
		\substmap(@(M_0, M_1),x,N) & \defeq
			@( \substmap(M_0,x,N), \substmap(M_1,x,N) ).
	\end{align*}
\end{exa}

\clearpage
\bibliographystyle{eptcs}
\bibliography{nominal-nu-mu-fics}

\begin{thebibliography}{10}
\providecommand{\bibitemdeclare}[2]{}
\providecommand{\surnamestart}{}
\providecommand{\surnameend}{}
\providecommand{\urlprefix}{Available at }
\providecommand{\url}[1]{\texttt{#1}}
\providecommand{\href}[2]{\texttt{#2}}
\providecommand{\urlalt}[2]{\href{#1}{#2}}
\providecommand{\doi}[1]{doi:\urlalt{https://doi.org/#1}{#1}}
\providecommand{\eprint}[1]{arXiv:\urlalt{https://arxiv.org/abs/#1}{#1}}
\providecommand{\bibinfo}[2]{#2}

\bibitemdeclare{article}{AdamekAl18}
\bibitem{AdamekAl18}
\bibinfo{author}{Jiří \surnamestart Adámek\surnameend},
  \bibinfo{author}{Stefan \surnamestart Milius\surnameend} \&
  \bibinfo{author}{Lawrence~S. \surnamestart Moss\surnameend}
  (\bibinfo{year}{2018}): \emph{\bibinfo{title}{Fixed points of functors}}.
\newblock {\slshape \bibinfo{journal}{Journal of Logical and Algebraic Methods
  in Programming}} \bibinfo{volume}{95}, pp. \bibinfo{pages}{41--81},
  \doi{10.1016/j.jlamp.2017.11.003}.

\bibitemdeclare{incollection}{AdamekMiVe09}
\bibitem{AdamekMiVe09}
\bibinfo{author}{Jiří \surnamestart Adámek\surnameend},
  \bibinfo{author}{Stefan \surnamestart Milius\surnameend} \&
  \bibinfo{author}{Jiří \surnamestart Velebil\surnameend}
  (\bibinfo{year}{2009}): \emph{\bibinfo{title}{Semantics of Higher-Order
  Recursion Schemes}}.
\newblock In: {\slshape \bibinfo{booktitle}{CALCO 2009: Algebra and Coalgebra
  in Computer Science}}, pp. \bibinfo{pages}{49--63},
  \doi{10.1007/978-3-642-03741-2_5}.

\bibitemdeclare{article}{AppelAl07}
\bibitem{AppelAl07}
\bibinfo{author}{Andrew~W. \surnamestart Appel\surnameend},
  \bibinfo{author}{Paul-André \surnamestart Melliès\surnameend},
  \bibinfo{author}{Christopher~D. \surnamestart Richards\surnameend} \&
  \bibinfo{author}{Jérôme \surnamestart Vouillon\surnameend}
  (\bibinfo{year}{2007}): \emph{\bibinfo{title}{A very modal model of a modern,
  major, general type system}}.
\newblock {\slshape \bibinfo{journal}{{ACM} {SIGPLAN} Notices}}
  \bibinfo{volume}{42}(\bibinfo{number}{1}), pp. \bibinfo{pages}{109--122},
  \doi{10.1145/1190215.1190235}.

\bibitemdeclare{phdthesis}{Arkor22}
\bibitem{Arkor22}
\bibinfo{author}{Nathanael \surnamestart Arkor\surnameend}
  (\bibinfo{year}{2022}): \emph{\bibinfo{title}{Monadic and Higher-Order
  Structure}}.
\newblock \bibinfo{type}{phdthesis}, \bibinfo{school}{University of Cambridge},
  \doi{10.17863/CAM.86347}.

\bibitemdeclare{article}{ArnoldNivat80}
\bibitem{ArnoldNivat80}
\bibinfo{author}{André \surnamestart Arnold\surnameend} \&
  \bibinfo{author}{Maurice \surnamestart Nivat\surnameend}
  (\bibinfo{year}{1980}): \emph{\bibinfo{title}{The metric space of infinite
  trees. Algebraic and topological properties}}.
\newblock {\slshape \bibinfo{journal}{Fundamenta Informaticae}}
  \bibinfo{volume}{3}(\bibinfo{number}{4}), pp. \bibinfo{pages}{445--475},
  \doi{10.3233/fi-1980-3405}.

\bibitemdeclare{book}{Barendregt84}
\bibitem{Barendregt84}
\bibinfo{author}{Henk~P. \surnamestart Barendregt\surnameend}
  (\bibinfo{year}{1984}): \emph{\bibinfo{title}{The Lambda Calculus. Its Syntax
  and Semantics}}, \bibinfo{edition}{2} edition.
\newblock {\slshape \bibinfo{series}{Studies in Logic and the Foundations of
  Mathematics}} \bibinfo{volume}{103}, \bibinfo{publisher}{Elsevier Science},
  \doi{10.1016/B978-0-444-87508-2.50006-X}.

\bibitemdeclare{article}{Barr93}
\bibitem{Barr93}
\bibinfo{author}{Michael \surnamestart Barr\surnameend} (\bibinfo{year}{1993}):
  \emph{\bibinfo{title}{Terminal coalgebras in well-founded set theory}}.
\newblock {\slshape \bibinfo{journal}{Theoretical Computer Science}}
  \bibinfo{volume}{114}(\bibinfo{number}{2}), pp. \bibinfo{pages}{299--315},
  \doi{10.1016/0304-3975(93)90076-6}.

\bibitemdeclare{incollection}{Berarducci96}
\bibitem{Berarducci96}
\bibinfo{author}{Alessandro \surnamestart Berarducci\surnameend}
  (\bibinfo{year}{1996}): \emph{\bibinfo{title}{Infinite λ-calculus and
  non-sensible models}}.
\newblock In: {\slshape \bibinfo{booktitle}{Logic and Algebra}},
  \bibinfo{publisher}{Routledge}, pp. \bibinfo{pages}{339--377},
  \doi{10.1201/9780203748671-17}.

\bibitemdeclare{article}{deBruijn72}
\bibitem{deBruijn72}
\bibinfo{author}{Nicolaas~Govert \surnamestart de~Bruijn\surnameend}
  (\bibinfo{year}{1972}): \emph{\bibinfo{title}{Lambda Calculus Notation with
  Nameless Dummies, a Tool for Automatic Formula Manipulation, with
  applications to the Church-Rosser Theorem}}.
\newblock {\slshape \bibinfo{journal}{Indagationes Mathematicæ}}
  \bibinfo{volume}{34}, \doi{10.1016/1385-7258(72)90034-0}.

\bibitemdeclare{phdthesis}{Cerda24}
\bibitem{Cerda24}
\bibinfo{author}{Rémy \surnamestart Cerda\surnameend} (\bibinfo{year}{2024}):
  \emph{\bibinfo{title}{Taylor Approximation for Infinitary λ-Calculi}}.
\newblock Ph.D. thesis, \bibinfo{school}{Aix-Marseille University}.

\bibitemdeclare{misc}{CerdaVaux23}
\bibitem{CerdaVaux23}
\bibinfo{author}{Rémy \surnamestart Cerda\surnameend} \&
  \bibinfo{author}{Lionel \surnamestart {Vaux Auclair}\surnameend}
  (\bibinfo{year}{2023}): \emph{\bibinfo{title}{Finitary Simulation of
  Infinitary β-Reduction via Taylor Expansion, and Applications}}.
\newblock \eprint{2211.05608}.

\bibitemdeclare{misc}{DalLago16}
\bibitem{DalLago16}
\bibinfo{author}{Ugo \surnamestart {Dal Lago}\surnameend}
  (\bibinfo{year}{2016}): \emph{\bibinfo{title}{Infinitary $\lambda$-Calculi
  from a Linear Perspective}}.
\newblock \eprint{1604.08248}.
\newblock \bibinfo{note}{Long version of a \emph{LICS} paper}.

\bibitemdeclare{inproceedings}{FioreAl99}
\bibitem{FioreAl99}
\bibinfo{author}{Marcelo \surnamestart Fiore\surnameend},
  \bibinfo{author}{Gordon \surnamestart Plotkin\surnameend} \&
  \bibinfo{author}{Daniele \surnamestart Turi\surnameend}
  (\bibinfo{year}{1999}): \emph{\bibinfo{title}{Abstract syntax and variable
  binding}}.
\newblock In: {\slshape \bibinfo{booktitle}{14th Symposium on Logic in Computer
  Science}}, \doi{10.1109/lics.1999.782615}.

\bibitemdeclare{article}{GabbayPitts02}
\bibitem{GabbayPitts02}
\bibinfo{author}{Murdoch~J. \surnamestart Gabbay\surnameend} \&
  \bibinfo{author}{Andrew~M. \surnamestart Pitts\surnameend}
  (\bibinfo{year}{2002}): \emph{\bibinfo{title}{A New Approach to Abstract
  Syntax with Variable Binding}}.
\newblock {\slshape \bibinfo{journal}{Formal Aspects of Computing}}
  (\bibinfo{number}{13}), pp. \bibinfo{pages}{341--363},
  \doi{10.1007/s001650200016}.

\bibitemdeclare{inproceedings}{HirschowitzEtAl22}
\bibitem{HirschowitzEtAl22}
\bibinfo{author}{André \surnamestart Hirschowitz\surnameend},
  \bibinfo{author}{Tom \surnamestart Hirschowitz\surnameend},
  \bibinfo{author}{Ambroise \surnamestart Lafont\surnameend} \&
  \bibinfo{author}{Marco \surnamestart Maggesi\surnameend}
  (\bibinfo{year}{2022}): \emph{\bibinfo{title}{Variable binding and
  substitution for (nameless) dummies}}.
\newblock In: {\slshape \bibinfo{booktitle}{FoSSaCS 2022: Foundations of
  Software Science and Computation Structures}}, pp. \bibinfo{pages}{389--408},
  \doi{10.1007/978-3-030-99253-8_20}.

\bibitemdeclare{article}{KennawayAl97}
\bibitem{KennawayAl97}
\bibinfo{author}{Richard \surnamestart Kennaway\surnameend},
  \bibinfo{author}{Jan~Willem \surnamestart Klop\surnameend},
  \bibinfo{author}{Ronan \surnamestart Sleep\surnameend} \&
  \bibinfo{author}{Fer-Jan \surnamestart de~Vries\surnameend}
  (\bibinfo{year}{1997}): \emph{\bibinfo{title}{Infinitary lambda calculus}}.
\newblock {\slshape \bibinfo{journal}{Theoretical Computer Science}}
  \bibinfo{volume}{175}(\bibinfo{number}{1}), pp. \bibinfo{pages}{93--125},
  \doi{10.1016/S0304-3975(96)00171-5}.

\bibitemdeclare{misc}{Kock09}
\bibitem{Kock09}
\bibinfo{author}{Joachim \surnamestart Kock\surnameend} (\bibinfo{year}{2009}):
  \emph{\bibinfo{title}{Notes on Polynomial Functors}}.
\newblock \urlprefix\url{https://mat.uab.cat/~kock/cat/polynomial.pdf}.

\bibitemdeclare{incollection}{KurzPetrisanAl12}
\bibitem{KurzPetrisanAl12}
\bibinfo{author}{Alexander \surnamestart Kurz\surnameend},
  \bibinfo{author}{Daniela \surnamestart Petrişan\surnameend},
  \bibinfo{author}{Paula \surnamestart Severi\surnameend} \&
  \bibinfo{author}{Fer-Jan \surnamestart {de Vries}\surnameend}
  (\bibinfo{year}{2012}): \emph{\bibinfo{title}{An Alpha-Corecursion Principle
  for the Infinitary Lambda Calculus}}.
\newblock In: {\slshape \bibinfo{booktitle}{CMCS 2012}},
  \bibinfo{publisher}{Springer}, pp. \bibinfo{pages}{130--149},
  \doi{10.1007/978-3-642-32784-1_8}.

\bibitemdeclare{article}{KurzPetrisanAl13}
\bibitem{KurzPetrisanAl13}
\bibinfo{author}{Alexander \surnamestart Kurz\surnameend},
  \bibinfo{author}{Daniela \surnamestart Petrişan\surnameend},
  \bibinfo{author}{Paula \surnamestart Severi\surnameend} \&
  \bibinfo{author}{Fer-Jan \surnamestart {de Vries}\surnameend}
  (\bibinfo{year}{2013}): \emph{\bibinfo{title}{Nominal Coalgebraic Data Types
  with Applications to Lambda Calculus}}.
\newblock {\slshape \bibinfo{journal}{Logical Methods in Computer Science}}
  \bibinfo{volume}{9}(\bibinfo{number}{4}), \doi{10.2168/lmcs-9(4:20)2013}.

\bibitemdeclare{article}{Lambek68}
\bibitem{Lambek68}
\bibinfo{author}{Joachim \surnamestart Lambek\surnameend}
  (\bibinfo{year}{1968}): \emph{\bibinfo{title}{A Fixpoint Theorem for complete
  Categories}}.
\newblock {\slshape \bibinfo{journal}{Mathematische Zeitschrift}}
  \bibinfo{volume}{103}, pp. \bibinfo{pages}{151--161},
  \doi{10.1007/BF01110627}.
\newblock \urlprefix\url{http://eudml.org/doc/170906}.

\bibitemdeclare{article}{LehmannSmyth81}
\bibitem{LehmannSmyth81}
\bibinfo{author}{Daniel~J. \surnamestart Lehmann\surnameend} \&
  \bibinfo{author}{Michael~B. \surnamestart Smyth\surnameend}
  (\bibinfo{year}{1981}): \emph{\bibinfo{title}{Algebraic specification of data
  types: A synthetic approach}}.
\newblock {\slshape \bibinfo{journal}{Mathematical Systems Theory}}
  \bibinfo{volume}{14}(\bibinfo{number}{1}), pp. \bibinfo{pages}{97--139},
  \doi{10.1007/bf01752392}.

\bibitemdeclare{article}{Moss01}
\bibitem{Moss01}
\bibinfo{author}{Lawrence~S. \surnamestart Moss\surnameend}
  (\bibinfo{year}{2001}): \emph{\bibinfo{title}{Parametric corecursion}}.
\newblock {\slshape \bibinfo{journal}{Theoretical Computer Science}}
  \bibinfo{volume}{260}(\bibinfo{number}{1–2}), pp.
  \bibinfo{pages}{139--163}, \doi{10.1016/s0304-3975(00)00126-2}.

\bibitemdeclare{unpublished}{Metayer03}
\bibitem{Metayer03}
\bibinfo{author}{François \surnamestart Métayer\surnameend}
  (\bibinfo{year}{2003}): \emph{\bibinfo{title}{Fixed points of functors}}.
\newblock \urlprefix\url{https://www.irif.fr/~metayer/PDF/fix.pdf}.

\bibitemdeclare{inproceedings}{Nakano00}
\bibitem{Nakano00}
\bibinfo{author}{Hiroshi \surnamestart Nakano\surnameend}
  (\bibinfo{year}{2000}): \emph{\bibinfo{title}{A modality for recursion}}.
\newblock In: {\slshape \bibinfo{booktitle}{Proceedings of the 15th Annual
  {IEEE} Symposium on Logic in Computer Science}},
  \doi{10.1109/lics.2000.855774}.

\bibitemdeclare{article}{Pitts06}
\bibitem{Pitts06}
\bibinfo{author}{Andrew~M. \surnamestart Pitts\surnameend}
  (\bibinfo{year}{2006}): \emph{\bibinfo{title}{Alpha-structural recursion and
  induction}}.
\newblock {\slshape \bibinfo{journal}{Journal of the ACM}}
  \bibinfo{volume}{53}(\bibinfo{number}{3}), pp. \bibinfo{pages}{459--506},
  \doi{10.1145/1147954.1147961}.

\bibitemdeclare{book}{Pitts13}
\bibitem{Pitts13}
\bibinfo{author}{Andrew~M. \surnamestart Pitts\surnameend}
  (\bibinfo{year}{2013}): \emph{\bibinfo{title}{Nominal Sets. Names and
  Symmetry in Computer Science}}.
\newblock \bibinfo{publisher}{Cambridge University Press},
  \doi{10.1017/CBO9781139084673}.

\bibitemdeclare{incollection}{Plotkin90}
\bibitem{Plotkin90}
\bibinfo{author}{Gordon \surnamestart Plotkin\surnameend}
  (\bibinfo{year}{1990}): \emph{\bibinfo{title}{An Illative Theory of
  Relations}}.
\newblock In: {\slshape \bibinfo{booktitle}{Situation Theory and Its
  Applications}}, \bibinfo{volume}{1}, \bibinfo{publisher}{CSLI}, pp.
  \bibinfo{pages}{133--146}.
\newblock \urlprefix\url{http://hdl.handle.net/1842/180}.

\bibitemdeclare{article}{Power07}
\bibitem{Power07}
\bibinfo{author}{John \surnamestart Power\surnameend} (\bibinfo{year}{2007}):
  \emph{\bibinfo{title}{Abstract Syntax: Substitution and Binders}}.
\newblock {\slshape \bibinfo{journal}{Electronic Notes in Theoretical Computer
  Science}} \bibinfo{volume}{173}, pp. \bibinfo{pages}{3--16},
  \doi{10.1016/j.entcs.2007.02.024}.

\end{thebibliography}

\end{document}